\documentclass{webofc}
\usepackage[varg]{txfonts}   
\usepackage{xcolor}
\usepackage[normalem]{ulem}  

\begin{document}
\title{Tension between implications from PREX-2 data and gravitational tidal response on dense matter equation of state}
%
%

\author{\firstname{Vivek Baruah} \lastname{Thapa}\inst{1,2}\fnsep\thanks{\email{thapa.1@iitj.ac.in}} \and
        \firstname{Monika} \lastname{Sinha}\inst{1}
}

\institute{Indian Institute of Technology Jodhpur, Jodhpur-342037, India 
\and
           National Institute of Physics and Nuclear Engineering (IFIN-HH), RO-077125, Bucharest, Romania 
          }

\abstract{%
  Recently an improved value of neutron skin thickness of $^{208}\text{Pb}$ was reported in Lead Radius EXperiment-2 (PREX-2) to be $R_{\text{skin}}=R_n - R_p=(0.283\pm 0.071)$ fm which corresponds to high estimations of nuclear symmetry energy ($E_{\text{sym}}$) and its slope ($L_{\text{sym}}$). The updated values of $E_{\text{sym}}$ and $L_{\text{sym}}$ commensurating to the neutron star observable estimations lie exterior to the astrophysical observed range. The higher values of $L_{\text{sym}}$ at $n_0$ deduced from recent PREX-2 data correlates to matter being easily deformable (yielding higher radius values) around intermediate matter densities leading to higher values of $\tilde{\Lambda}$ creating a tension between the terrestrial and astrophysical observations. In this study, we exploit this tension to constrain the $\Delta$-scalar meson coupling parameter space.
}
\maketitle
\section{Introduction} \label{intro}

The neutron stars (NSs), highly compact stars, contain densest matter inside them. The matter density inside the NS core varies in the range of a few times nuclear matter density \cite{1996cost.book.....G}.  
Most of the models of highly dense matter are formulated to reproduce the experimentally obtained range of the saturation property parameters, viz., nuclear saturation density ($n_0$), saturation energy ($E_0$), incompressibility ($K_0$), symmetry energy ($E_{\rm{sym}}$), its slope with density ($L_{\rm{sym}}$), curvature of symmetry energy ($K_{\rm{sym}}$), and effective nucleonic Dirac mass ($m^*_N$).
There is recent update in obtaining the experimental values of $E_{\rm{sym}}$ and its density dependence from the nuclear physics experiment. The Lead Radius EXperiment (PREX) collaboration reported its first results (PREX-1) \cite{PhysRevLett.108.112502} with $R_{\text{skin}}=R_n - R_p=0.33^{+0.16}_{-0.18}$ fm and corresponding $L_{\text{sym}}(n_0)$ based on strong correlations to be in the range ($35-265$) MeV. 
Later on the ranges of $E_{\text{sym}}$ and $L_{\text{sym}}$ based on experimental data from finite nuclei and heavy-ion collisions (HICs) with different microscopic model calculations were estimated to be $28.5-34.9$ MeV and $30.6-86.8$ MeV respectively \cite{2017RvMP...89a5007O}.
A latest compilation has been reported in ref.-\cite{PhysRevLett.125.202702} for the isospin properties and that to be in the range of ($30.6-32.8$) MeV, ($55.7-63.9$) MeV for $E_{\text{sym}}$ and $L_{\text{sym}}$ at nuclear saturation density ($n_0$) respectively. But very recently an improved value of $^{208}\text{Pb}$ was reported in PREX-2 as $R_{\text{skin}}=(0.283\pm 0.071)$ fm \cite{PhysRevLett.126.172502} with $\sim 1 \%$ precision.
This however leads to an estimation of $E_{\rm{sym}}$ and $L_{\rm{sym}}$ at $n_0$ to be in the ranges ($38.1 \pm 4.7$) MeV, ($106\pm 37$) MeV respectively ($1\sigma$ interval) with correlation coefficient as 0.978 \cite{PhysRevLett.126.172503}. 
%
%
%
The dense matter model should be tested and constrained with the recent astrophysical observations from NSs along with the nuclear physics experimental data.
In recent years, the pursuit to determine and constrain the dense matter equation of state (EOS) can be drawn from the massive NS ($M\geq 2~M_{\odot}$) observations \cite{Arzoumanian_2018, 2013Sci...340..448A, 2021arXiv210400880F}, radius measurements of NS candidates from NICER (Neutron star Interior Composition ExplorER) space mission \cite{2019ApJ...887L..24M,2019ApJ...887L..21R, 2020PhRvD.101l3007L, 2021arXiv210506980R, 2021arXiv210506979M} as well as tidal response from gravitational-wave (GW) events \cite{LIGO_Virgo2017c, LIGO_Virgo2018a, PhysRevX.9.011001, 2020ApJ...892L...3A} by the LIGO-Virgo Collaboration. 
Very recently another pulsar, namely PSR J$0952-0607$ has been observed with mass reported in the range $2.18-2.52~M_\odot$ at $1-\sigma$ confidence interval \cite{2022arXiv220705124R}.
The large lower bound of maximum NS mass leads to consider the appearance of heavier baryons inside the inner core of the NS \cite{Drago_PRC_2014, Kolomeitsev_NPA_2017, Li2020PhRvD, particles3040043, 2021MNRAS.507.2991T, PhysRevC.105.015802}. Also the constrained of matter to be soft at intermediate density opens the possibility of $\Delta$ resonances to appear with the increase of density.

In present work we model the dense matter inside NS making compatible with recently updated ranges of $E_{\rm{sym}}$ and its density dependence as well as recent astrophysical observations within the covariant density functional (CDF) model of dense matter.
(\textit{Conventions}: We implement the natural units $G=\hbar=c=1$ throughout the work)

\section{Hadronic CDF model} \label{sec-1}

In this study, to compute the dense matter EOS we implement CDF framework model. In this model scheme, the coupling constants are based on the fact to reproduce the experimental quantities known at nuclear saturation. For this work, we consider the entire baryon octet ($b\equiv N,Y$) along with the spin-3/2 $\Delta$-resonances in the composition of dense matter. Leptons ($e^-,\mu^-$) are brought into the picture to maintain the beta-equilibrium condition.
We consider the effective interactions between the baryons are mediated via the isoscalar-scalar meson $\sigma$, isoscalar-vector mesons $\omega$, $\phi$ and the isovector-vector $\rho$-meson.
The hidden strangeness $\phi$-meson mediates the repulsive interactions between the strange baryons only.
The total Lagrangian density in this formalism is given by \cite{1996cost.book.....G}
\begin{equation}\label{eqn.1}
\begin{aligned}
\mathcal{L} & = \sum_{b} \bar{\psi}_b(i\gamma_{\mu} D^{\mu}_{(b)} - m^{*}_b) \psi_b + \sum_{l} \bar{\psi}_l (i\gamma_{\mu} \partial^{\mu} - m_l)\psi_l 
 + \sum_{\Delta} \bar{\psi}_{\Delta \nu}(i\gamma_{\mu} D^{\mu}_{(\Delta)} - m^{*}_{\Delta}) \psi^{\nu}_{\Delta} + \frac{1}{2}(\partial_{\mu}\sigma\partial^{\mu}\sigma \\
 & - m_{\sigma}^2 \sigma^2) -  \frac{1}{4}\omega_{\mu\nu}\omega^{\mu\nu} + \frac{1}{2}m_{\omega}^2\omega_{\mu}\omega^{\mu} - \frac{1}{4}\boldsymbol{\rho}_{\mu\nu} \cdot \boldsymbol{\rho}^{\mu\nu} + \frac{1}{2}m_{\rho}^2\boldsymbol{\rho}_{\mu} \cdot \boldsymbol{\rho}^{\mu}  - \frac{1}{4}\phi_{\mu\nu}\phi^{\mu\nu} + \frac{1}{2}m_{\phi}^2\phi_{\mu}\phi^{\mu} - \text{U}(\sigma),
\end{aligned}
\end{equation}
where the covariant derivative is given by $D_{\mu (j)} = \partial_\mu + ig_{\omega j} \omega_\mu + ig_{\rho j} \boldsymbol{\tau}_{j3} \cdot \boldsymbol{\rho}_{\mu} + ig_{\phi j} \phi_\mu$ with $j$ denoting the baryon particle spectrum. The baryon octet and $\Delta$ baryon Schwinger-Rarita fields along with their respective masses are represented by $\psi_b,~m_b$ and $\psi_\Delta,~m_{\Delta}$ respectively.
$\omega_{\mu \nu}$, $\boldsymbol{\rho}_{\mu \nu}$ and $\phi_{\mu \nu}$ are the anti-symmetric field tensors corresponding to vector meson fields. 
To reproduce the nuclear matter incompressibility at $n_0$, the self-interaction of $\sigma$ meson is included \cite{1996cost.book.....G} by the term $U(\sigma)=(1/3)g_2\sigma^3 + (1/4)g_3\sigma^4$ with $g_2$, $g_3$ denoting the self-interaction coefficients (non-linear (NL) model).
Alternate to the NL model is the density dependent (DD) coupling model. In latter model, the incompressibility of nuclear matter at $n_0$ can be reproduced without self-interaction term but considering the coupling parameters density dependent.
In the density-dependent approach, a rearrangement term is necessary to maintain the thermodynamic consistency contributing to matter pressure explicitly via the chemical potential and is given by $\Sigma^{r} = \sum_{b} \left[ \frac{\partial g_{\omega b}}{\partial n}\omega_{0}n_{b} - \frac{\partial g_{\sigma b}}{\partial n} \sigma n_{b}^s + \frac{\partial g_{\rho b}}{\partial n} \rho_{03} \boldsymbol{\tau}_{b3} n_{b} + \frac{\partial g_{\phi b}}{\partial n}\phi_{0}n_{b} \right] + \sum_{\Delta} (\psi_b \longrightarrow \psi_{\Delta}^{\nu})$.
The scalar couplings with the $\Lambda$, $\Sigma$-hyperons are determined from the hypernuclear binding energy fits corresponding to $U_{\Lambda}^N(n_0)=-30$ MeV and $U_{\Sigma}^N(n_0)=+30$ MeV \cite{RevModPhys.88.035004}.
We have considered the $\sigma-\Xi$ coupling to be $U_{\Xi}^N (n_0)= -20$ MeV in symmetric nuclear matter reported recently in Ref.-\cite{2021arXiv210400421F}.
As for the vector meson couplings with the hyperons, they are incorporated according to SU(6) symmetry \cite{SCHAFFNER199435}.
As the $\Delta$-baryons are resonant states of nucleons, they do not couple with isoscalar-vector $\phi$-meson. Due to insufficient knowledge on meson-$\Delta$ couplings in nuclear matter, we treat these couplings as parameters.
The $\Delta$ isoscalar meson coupling gap is constrained in Ref.-\cite{Drago_PRC_2014} to be the range $0 \leq R_{\sigma \Delta}-R_{\omega \Delta} \leq 0.2$ where $R_{i \Delta}=g_{i \Delta}/g_{i N}$ and $i=\sigma,~\omega,~\rho$-mesons.
In this study, we further constrain this parameter space on the basis of reconciling the tension between experimental nuclear and astrophysical observations.
Here, we consider the isoscalar-vector and isovector-vector $\Delta$-couplings to be $R_{\omega \Delta}=1.1$, $R_{\rho \Delta}=1.0$ respectively.
The scalar $\Delta$-coupling parameter space is varied in the range $1.1 \leqslant R_{\sigma \Delta} \leqslant 1.3$.

\section{Results $\&$ Discussion} \label{sec:results}

This section discusses the implications of exotic particles in reconciling the tension between the nuclear physics and astrophysical constraints.
Here, we report the numerical results of various dense matter EOSs corresponding to variation in $R_{\sigma \Delta}-R_{\omega \Delta}$ coupling parameter space and isospin dependent saturation property.

\begin{figure*}[t!]
    \centering
    \sidecaption
    \includegraphics[width=7.0cm, keepaspectratio]{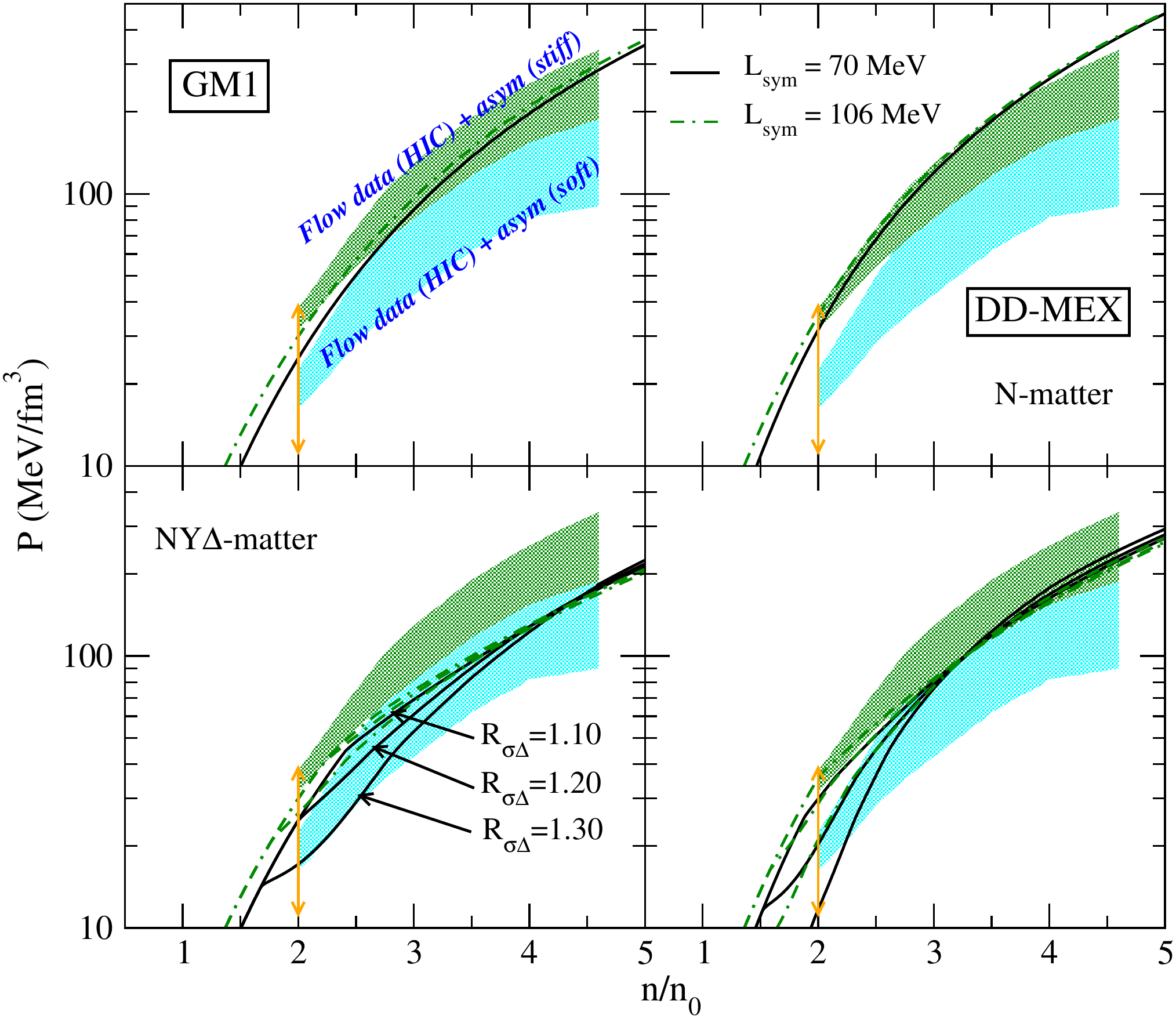}
    \caption{The matter pressure as a function of baryon number density for the nucleonic (upper panels), $\Delta$-admixed hypernuclear (lower panels) dense matter EOSs considering variation in $L_{\text{sym}}(n_0)$ and scalar meson-$\Delta$ coupling values. Re-calibrated $L_{\text{sym}}(n_0)=70,~106$ MeV EOSs are designated via the solid and dot-dashed curves respectively. The vertical bound at $2n_0$ is depicted from GW170817 event data via interpolation \cite{LIGO_Virgo2018a}. The shaded regions from density range $(2-4.5)~n_0$ depict the flow data from HIC and modelled with stiffer as well as softer EOSs \cite{2002Sci...298.1592D}.}
    \label{fig.001}
\end{figure*}
\begin{figure}[h!]
    \centering
    \sidecaption
    \includegraphics[width=7.0cm, keepaspectratio]{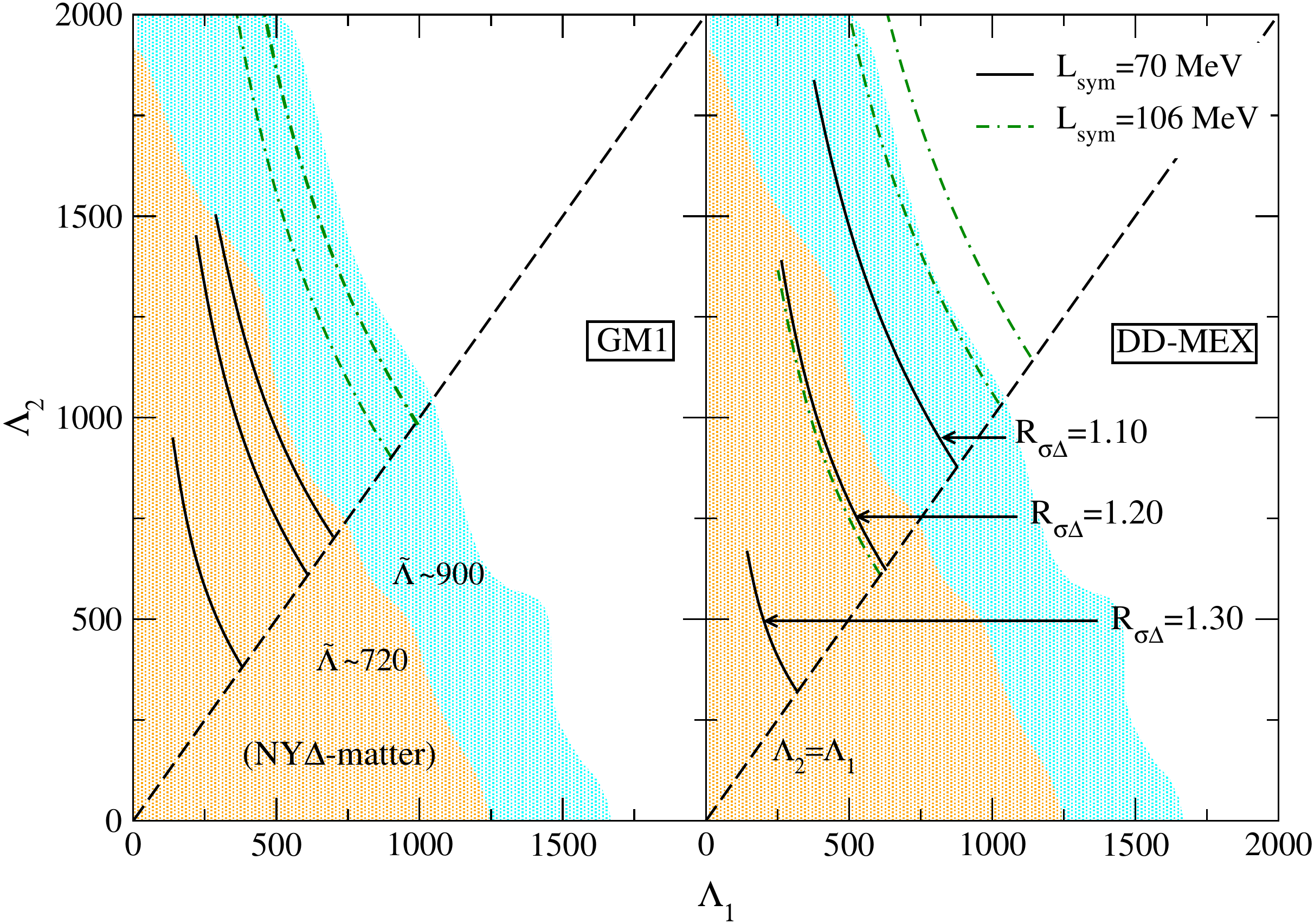}
    \caption{Tidal deformability parameters corresponding to the primary and secondary components involved in GW170817 event for various isospin dependent coupling in accordance to PREX-2 and alterations in $R_{\sigma \Delta}-R_{\omega \Delta}$. Here, we have considered the chirp mass to be $\mathcal{M}=1.186~M_{\odot}$. The shaded regions denote the tidal deformability upper bounds of $\tilde{\Lambda}\sim900$ \cite{LIGO_Virgo2017c} (cyan) and $\tilde{\Lambda}\sim720$ \cite{PhysRevX.9.011001} (orange).}
    \label{fig.002}
\end{figure}
The pressure variation with baryon number density for the different EOS models are provided in fig.-\ref{fig.001}.
It can be seen from the figure that the re-calibrated EOSs obtained via exploiting the $L_{\text{sym}}$ and $\sigma-\Delta$ coupling parameter space for both the parametrization models lie within the constraints from GW170817 as well as the HIC data.
Here, it is also evident that from the figure that considering the upper bound on $\sigma-\Delta$ coupling, $R_{\sigma \Delta}=1.3$ satisfies the constraints from terrestrial as well as astrophysical observations.
With further attractive optical potential for $\Delta$-resonances in nuclear matter, the EOS models tend to be softer at the lower matter density regimes as evident from the figure.
This result is in consistent with Ref.-\cite{PhysRevC.105.015802}.
The EOSs considered in this work fulfill the thermodynamic stability condition in addition to providing non-vanishing effective Dirac nucleon mass as pointed in Ref.-\cite{2022arXiv220602935M}.

The tidal responses of both the components involved in GW170817 event are evaluated based on dense matter EOSs with variation in symmetry energy slope and scalar $\Delta$-meson couplings.
From the fig.-\ref{fig.002}, it can be seen that with higher values of $L_{\text{sym}}$, the tidal responses also gradually increases. This is because of the increase in radius of intermediate mass NSs and $\tilde{\Lambda}$ is proportional to fifth-power of NS radius.
And with higher values of $\sigma-\Delta$ coupling corresponding to further attractive potential, the tidal deformability is observed to decrease as invoking $\Delta^-$ at early matter densities induces more compactness.
With the strict constraint of $\tilde{\Lambda}\sim 720$ obtained from recent reanalysis of GW170817 event data, it can be inferred that attractive potentials of $\Delta$-resonances in dense matter is required to satisfy this constraint.

\begin{figure}[h!]
    \centering
    \sidecaption
    \includegraphics[width=7.5cm, keepaspectratio]{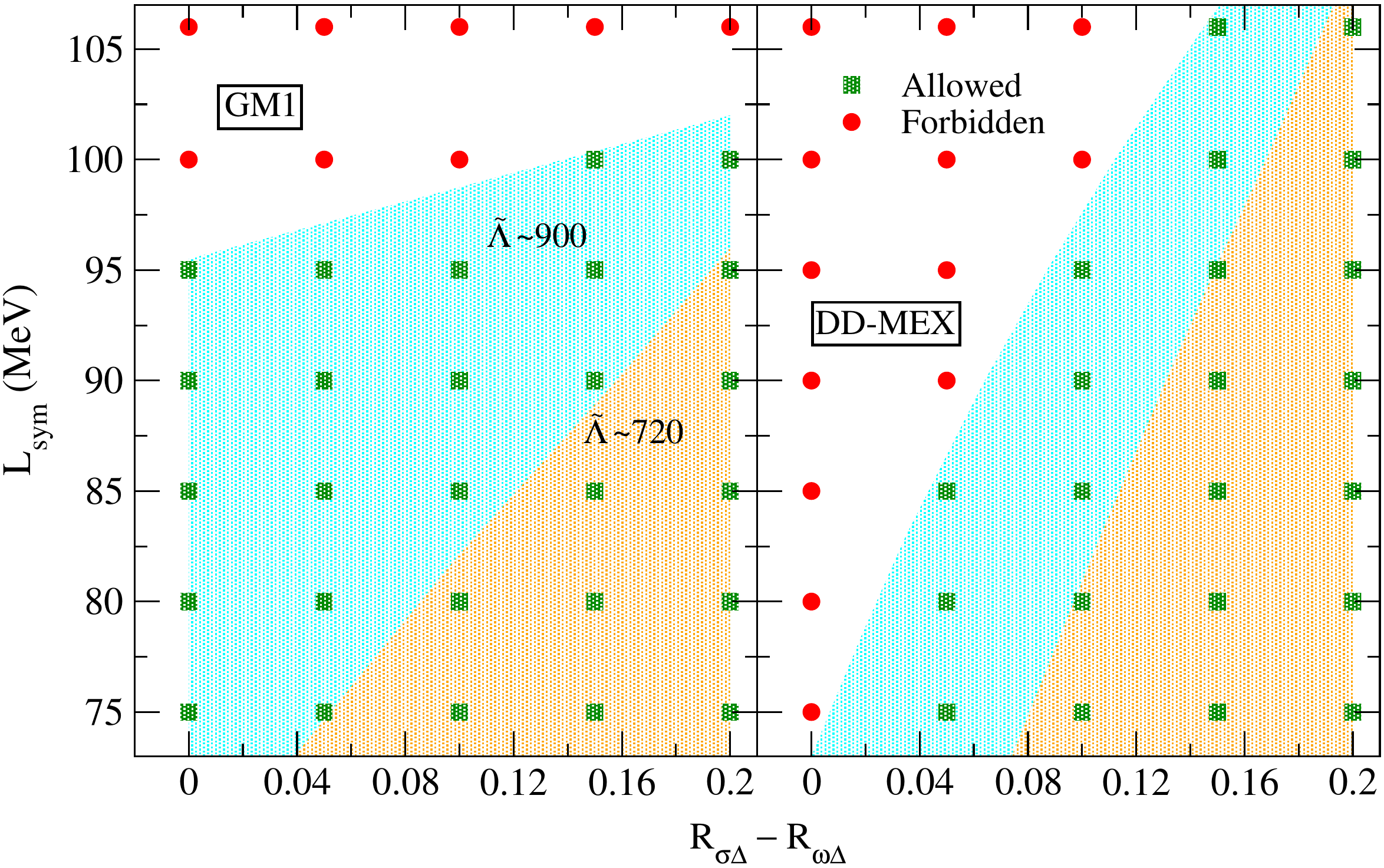}
    \caption{Coupling parameter space range of $R_{\sigma \Delta}-R_{\omega \Delta}$ for which constraints from nuclear physics experiments as well as astrophysical constraints are fulfilled. Here, we have varied $R_{\sigma \Delta}$ in the range of $[1.10-1.30]$. The shaded regions (cyan $\longrightarrow~\tilde{\Lambda}\sim 900$ and orange $\longrightarrow~\tilde{\Lambda}\sim 720$) are the allowed parameter space which fulfill both the terrestrial and astrophysical constraints.}
    \label{fig.003}
\end{figure}
The variation of $\Delta$-coupling parameters with $L_{\text{sym}}$ in dense matter EOS is shown in fig.-\ref{fig.003}.
This figure provides the allowed range of $R_{\sigma \Delta}$ keeping $R_{\omega \Delta}=1.10$ and $R_{\rho \Delta}=1.0$ with changing $L_{\text{sym}}$ scenarios.
It can be seen that higher $L_{\text{sym}}$ values demand the $\Delta$ potential in dense matter to be attractive in nature.
In case of DD-MEX parameterization, for $L_{\text{sym}}\leq 80$ MeV, the $\Delta$ coupling parameter space is unconstrained and all values of $R_{\sigma \Delta}-R_{\omega \Delta}$ are admissible following the former $\tilde{\Lambda}\sim 900$ constraint.
While in case of GM1 coupling set, this $\tilde{\Lambda}$ constraint bounds the $R_{\sigma \Delta}-R_{\omega \Delta}$ parameter set for $L_{\text{sym}}\leq 95$ MeV.
Following strict upper bound of $\tilde{\Lambda}\sim 720$, the parameter space is further constrained leaving the coupling value of $R_{\sigma \Delta}=1.30$ in case of DD-MEX set.
However, it is noteworthy to mention that another analysis of the GW170817 event data \cite{2021PhRvD.103l4015G} suggest much higher tidal deformability estimations $\sim 1000$ which would definitely put loose constraints on the poorly known $\Delta-$coupling values.

In another recent experiment to measure $R_{\text{skin}}$ in $^{48}$Ca-isotope (CREX) \cite{PhysRevLett.129.042501}, the same has been reported to be $(0.121 \pm 0.026)$ fm which is in disagreement with PREX results.
CREX findings indicate the symmetry energy to be low consequently leading to more compact NSs.
This indicates a need to further investigate the limitations of present dense matter models so that both of the findings can be accomodated which is beyond the scope of this work.

%
%
%

\bibliography{references}
\end{document}